\documentclass{mem}
\usepackage{natbib}\usepackage{txfonts}\usepackage{balance}
\usepackage{graphicx}
\usepackage[a4paper,breaklinks,dvipdfm]{hyperref}
\idline{75}{282}
\begin{document}
\def\teff{$T\rm_{eff }$}
\def\kms{$\mathrm {km s}^{-1}$}

\title{
On the PDS of GRB light curves
}

   \subtitle{}

\author{
S. \,Bo\c{c}i\inst{1} 
\and M. \, Hafizi\inst{1}
          }

\institute{
University of Tirana --
Faculty of Natural Sciences,
Tirana, Albania
\email{sonila.boci@fshn.edu.al,}
\email{mimoza.hafizi@fshn.edu.al}
}

\authorrunning{Bo\c{c}i, Hafizi}

\titlerunning{On the PDS of GRB light curves}

\abstract{In spite of the complicated behavior in the time
domain, long GRBs show a simpler behavior in the
Fourier domain of frequencies, represented by power density spectra, PDS. Recently, there are some relations found between GRBs properties and PDS
parameters, modeled by power-laws. Among them, the correlation between peak energy $E_{peak}$ and PDS slope $\alpha$ shows a clear evidence. In this work we try to understand the origin of this correlation, making use of synthetic pulses. We find some preliminary evidences that $E_{peak}-\alpha$ relation can be seen as a new confirmation of the empiric relations $E_{peak}-L$ and $t_{p}-L$ for GRBs.   

}
\maketitle{}

\section{Introduction}

The light curves of GRBs typically have many random peaks, diverse structures which appear to be the result of a complex distribution of several pulses. Burst pulses are commonly described by a fast rise exponential-decay (FRED) shape, although the decay is not strictly exponential \citep{Norris}:
\begin{equation}
    c(t)=A\lambda\exp[-\frac{\tau}{t-t_s}-\frac{t-t_s}{\tau_2}]
\end{equation}
or power-law shape \citep{Kocevski}:
\begin{equation}\label{eq:2}
    c(t)=F_m(\frac{t}{t_m})^r[\frac{d}{d+r}+\frac{r}{d+r}(\frac{t}{t_m})^{r+1}]^{-\frac{r+d}{r+1}}
\end{equation}

Analysis has shown broad log-normal distribution
in time duration, not only among different bursts,
but also within a single burst.
In spite of extensive studies, this temporal
behavior is not still understood. 
Light curve analysis can be a powerful tool to
shed light on the still obscure physics and
geometry of the prompt emission of GRBs. It can
provide insights into the size, the distance of the
dissipation region and the radiation processes.

In spite of the complicated behavior in the time
domain, long GRBs show a simpler behavior in the
Fourier domain of frequencies $c(t)\rightarrow
C(f)=\int_{-\infty}^{\infty}c(t)e^{2\pi i ft}dt$.

The power density spectrum PDS is defined as
$P_f=C_fC_f^*$.

In the case of observed curves with $N$ discrete
data $c_m$ (time sequence), the discrete Fourier transform DFT is estimated by:

$C_k=\sum_{m=0}^{N-1}c_m e^{2\pi imn/N}$,
$P(f_k)=|C_k|^2$.

\section{Computing PDS}

One frequently used way of estimating PDS is the periodogram, given by

$P(f_k)=\frac{1}{N^2}[|C_k|^2+|C_{N-k}|^2]$, $k=1,2,...,\frac{N}{2}$.

$P(f_k)$ is considered as the average of $P(f)$ over
a narrow window function, centered on $f_k$. This
window function would naturally be
$W(f)=\frac{1}{N^2}[\frac{\sin \pi f}{\sin \pi
f/N}]$, the Fourier transform of the rectangular
function. This window function is not zero outside the
corresponding frequency interval, so the periodogram
estimate is influenced from other frequencies
outside the interval, technically speaking leaks
from one frequency to another. The correction of the
leakage is called data windowing.
Instead of the rectangular function, one chooses a
window function that changes more gradually from
zero to its maximum and then back to zero.
In our calculations, we use Bartlett function, but there are several different ones \citep{Press}.

Another question of the periodogram is the value of the standard deviation $\sigma$, which is $100\%$
of the estimate, independent on the number of data $N$.
There are several techniques for reducing the
standard deviation. The technique we make
use is to partition the original $N$ data ($N=2MK$)
in $2M$ segments, each of $K$ points. We choose a sequence of $2M$ points, one point from
each segment and repeat the procedure $K$ times,
for $K$ consecutive sampled points. Each sequence
is separately Fourier transformed to produce a
periodogram estimate \citep{Press}. 
Finally, the $K$ periodogram estimates are averaged
at each frequency. This final averaging reduces
the standard deviation of the estimate by $\sqrt{K}$.

There are other ways used for reducing the
deviation of a single PDS, one of them by using
Monte-Carlo simulations of synthetic GRBs around a
real one \citep{Ukwatta}. 

\section{Average and individual PDS}

\subsection{Average PDS}

The procedure of averaging consists in summing up
the PDS of individual bursts after some
normalization and dividing the result by the
number of bursts in the sample. The distribution of the individual $P_{f}$ around $<P_{f}>$ follows a standard exponential law, so the amplitude of
fluctuations in $<P_{f}>$ is given by: $\frac{<\Delta P_f>}{<P_f>}\sim N^{-1/2}$, $N$ the number of bursts
in the sample \citep{Beloborodov}. 

There are different ways of normalization: 

-The light curves are normalized to their peak 
 \citep{Beloborodov}; 
 
-The averaging is performed inside a sole group of
variability, taking into account also a kind of
pseudoredshift, obtained through empirical relations \citep{Lazzati};

-The averaging is performed inside sub-classes of
GRBs found based on the auto-correlation function
and considering the measured redshift \citep{Borgonovo}. 

The procedure of averaging follows the
conviction that
different GRBs are many realizations of a unique
stochastic process, giving rise to the variety of the observed profiles.

 The average PDS are modeled by a power law
 \citep{Beloborodov} or smoothly broken power-law
 \citep{Guidorzi}, extending over two frequency
 decades, from about $10^{-2}$ to 1 or 2 Hz. The
 power law index lies in the range $1.5-2$ and the
 break around $1-2$ Hz. 

\subsection{Individual PDS}

While the average PDS over a large number of GRBs
exhibits small fluctuations and is easier to
characterize, it provides no clues on the variety of
properties of individual GRBs. On the other hand,
the wide variety of light curves exhibited by GRBs
would potentially be indicative of different
emission and scattering processes.

The key point of studying individual versus
averaged PDS is that one can investigate the
possible connection between PDS and GRBs key
properties of prompt emission, such as peak energy
or the isotropic-equivalent radiated energy \citep {Dichiara}.

As mentioned above, one way of estimating individual PDS is by
simulating $N$ light curves around the real one,
by using Monte Carlo technique. The standard deviation
of simulated PDS is found and the individual PDS is
calculated inside this uncertainty \citep{Ukwatta}. This procedure follows as well the idea of a unique stochastic process inside a GRB.  

Otherwise, there are authors who calculate
only a single PDS over the entire observation duration \citep{Dichiara}, \citep{Guidorzi 2016}. Each GRB time profile is considered individually as the unique sample of a unique stochastic process, which is different from other GRBs. PDS and uncertainties at each frequency are calculated assuming Leahy normalization.

To fit an individual PDS, two models are used:

-the simpler one is a mere power-law plus the
white-noise constant, $S_{PL}(f)=Af^{-\alpha} +B$, $A$
the normalization constant,  $\alpha$ the power law
index and $B$ the white noise level;

-bent power law model,
$S_{PL}(f)=A[1+(\frac{f}{f_b})^\alpha]^{-1}+B$.

The power-law index lies in the range $1.5-4$, with
some exceptions to 6.

\section{Some relations}

There are some relations found between PDS
parameters and GRBs properties. The first group of
relations concerns the variability of GRBs and the
second one the GRBs energy.

In the first group we could mention a correlation
found between the dominant frequency and the
variability measure and another correlation between
the break frequency and the variability measure \citep{Lazzati}. The number of pulses also seems to
anti-correlate with the slope $\alpha$.  

\citet{Ukwatta} found that the redshift corrected
threshold frequency is positively correlated with the
isotropic peak luminosity. 

An interesting correlation is found between peak
energy and the PDS power law index $\alpha$ of individual GRBs, $E_{peak}-\alpha$ relation \citep{Dichiara}.

\section{A synthetic $E_{peak}-\alpha$ relation}

 We try to reproduce the above mentioned
 $E_{peak}-\alpha$ relation by 
 generating synthetic pulses with different luminosity and by combining some
 empiric relations for GRBs:
 
 -Luminosity - duration relation \citep{Hakkila}:
 
$L=3.4\times 10^{52} t_{p}^{-0.85}$,

with $t_{p}$ the pulse duration between the two
$Ae^{-3}$ intensity points, where $A$ is the intensity at the maximum of the pulse.

-Peak energy-luminosity $E_{peak}-L$ relation
\citep{Ghirlanda}:

$E_{peak}=380 (\frac{L}{1.6\times 10^{52}})^{0.43}$

 -A correlation between the rise time and pulse width measured by FWHM \citep{Kocevski}:
 
$t_{m}=0.323t_{p}(1+z)^{0.6}$, 

where
$t_{m}$ is the pulse rise time and
$t_{pobs}=t_{p}(1+z)^{0.6}$  is the duration of the
pulse to the observer.

Assuming z=1 and normalized Band spectrum, we calculate pulse peak photon flux to the observer as \citep{Boci}:

     $P=\frac{L}{4\pi D_{L}^{2}\Phi_{0} E_{peak}^{obs}} \int_{\frac{E_{A}}{E_{peak}^{ob s}}} ^ {\frac{E_{B}}{E_{peak}^{obs}}} B(x) dx $
     
Bursts pulses are described by equation \ref{eq:2}, with  $r=1.49$ and $d=2.39$. 

We calculate the PDS for each pulse and find the slope in logarithmic scale for the interval of frequencies between $(0.3-1)$ Hz. 
\begin{figure}[]
\resizebox{\hsize}{!}{\includegraphics[clip=true]{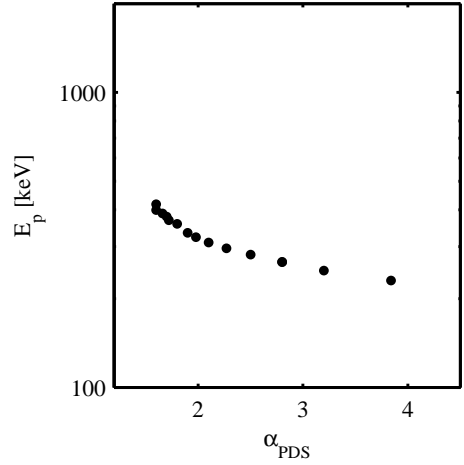}}
\caption{
\footnotesize
 $E_{peak}-\alpha$ diagram for the synthetic pulses with different luminosity. The slope $\alpha$ for each corresponding PDS is estimated by a power-law model, whereas $E_{peak}$ is found based on Ghirlanda $E_{peak}-L$ empiric relation.}
\label{Fig.1}
\end{figure}
The results are plotted in Fig.\ref{Fig.1}.

\section{Conclusions}

Although the considered sample is small, it seems that pulses with larger peak energy $E_{peak}$ exhibit lower power-law index $\alpha$, a trend similar to that found by \citep{Dichiara}. This result is only a preliminary one. We are performing a more systematic work by producing a larger sample of GRB's pulses. 

We believe that the confirmation of the $E_{peak}-\alpha$ correlation, making use of the synthetic pulses, 
can be seen as another evidence for empiric $E_{peak}-L$ and $t_{p}-L$ relations, thought to be important characteristics for 
using GRBs as tools in Cosmology, for scrutinizing the dark ages of the Universe, which is one of the 
important goals for the THESEUS mission \citep{amati17}. THESEUS will be very helpful to enlarge the sample of 
GRB ligtcurves for improving the correlations mentioned in this work.

\begin{acknowledgements}
We thank Lorenzo Amati and Cristiano Guidorzi for useful discussions on PDS, before and during Theseus workshop, Napoli, October 2017.
\end{acknowledgements}

\bibliographystyle{aa}

\end{document}